\documentclass[aps,prapplied,reprint,superscriptaddress,longbibliography]{revtex4-2}

\usepackage{soul}
\usepackage{graphicx}
\usepackage{amsmath,amssymb,amsfonts}
\usepackage{enumitem}
\usepackage{xcolor}
\usepackage{tikz,tkz-euclide}
\usepackage[colorlinks=true,citecolor=blue,linkcolor=blue,urlcolor=blue]{hyperref}
\usepackage[capitalise]{cleveref}
\usepackage{tikz,circuitikz, comment}
\usetikzlibrary{shadows,shadings,decorations.pathreplacing,calligraphy}
\usepackage{ifthen}
\usepackage{natbib}
\usepackage{braket}
\usepackage[caption=false]{subfig}

\newcommand{\w}{\omega}

\begin{document}

\title{
Suppressing chaos with mixed superconducting qubit devices}

\author{Ben Blain}
\affiliation{Quantum Research Center, Technology Innovation Institute, Abu Dhabi 9639, UAE}
\affiliation{School of Physics and Astronomy, University of Kent, Canterbury CT2 7NH, United Kingdom}

\author{Giampiero Marchegiani}
\email{giampiero.marchegiani@tii.ae}
\affiliation{Quantum Research Center, Technology Innovation Institute, Abu Dhabi 9639, UAE}

\author{Luigi~Amico}
\affiliation{Quantum Research Center, Technology Innovation Institute, Abu Dhabi 9639, UAE}
\affiliation{Dipartimento di Fisica e Astronomia, Via S. Sofia 64, 95123 Catania, Italy}
\affiliation{INFN-Sezione di Catania, Via S. Sofia 64, 95127 Catania, Italy}

\author{Gianluigi Catelani}
\affiliation{JARA Institute for Quantum Information (PGI-11),Forschungszentrum J\"ulich, 52425 J\"ulich, Germany}
\affiliation{Quantum Research Center, Technology Innovation Institute, Abu Dhabi 9639, UAE}

\begin{abstract}

In quantum information processing, a tension between two different tasks occurs: while qubits' states can be preserved by isolating them, quantum gates can be realized only through qubit-qubit interactions.  
In arrays of qubits, weak coupling leads to states being spatially localized and strong coupling to delocalized states.
Here, we study the average energy level spacing and the relative entropy of the  distribution of the level spacings  (Kullback-Leibler divergence from Poisson and Gaussian Orthogonal Ensemble) to analyze the crossover between localized and delocalized (chaotic) regimes in linear arrays of superconducting qubits.
We consider both transmons as well as capacitively shunted flux qubits, which enables us to tune the qubit anharmonicity. 
Arrays with uniform anharmonicity, comprising only transmons or flux qubits, display remarkably similar dependencies of 
level statistics on the coupling strength. 
In systems with alternating anharmonicity, for typical disorder in the qubit frequencies the localized regime is found to be more resilient to the increase in qubit-qubit coupling strength in comparison to arrays with a single qubit type. Our results, which we also confirm using generalized Bose-Hubbard models, support designing devices that incorporate different qubit types to achieve higher performances.
\end{abstract}

\maketitle

\section{Introduction} 
Superconducting qubits provide a feasible platform  for  quantum computing and simulation purposes~\cite{SCqubit_review}. 
For computation, fast quantum gates can be attained if the qubits are coupled to each other with sufficient strength~\cite{Sheldon2016,Wallraff2021Circuit,10.1063/1.5089550}. On the other hand, qubits need to be sufficiently isolated to minimize the effects of residual interactions that negatively affect information processing protocols. As such, studying this type of trade-off is a  problem of central importance which is attracting considerable attention in the field~\cite{Berke2022,Silveri2022natcomm,niko2024}.
It has indeed been argued that qubit arrays for quantum computation should be in a localized regime that can be achieved through a certain amount of spread (disorder) in qubit transition frequencies~\cite{Berke2022}.
Subsequent works both showed that quasiperiodic parameter modulation can be more effective than random disorder in keeping the system in the localized regimes~\cite{DiVincenzo2} and considered the connection to classical chaos in coupled nonlinear oscillators~\cite{DiVincenzo3}. Fixed qubit-qubit couplings lead to so-called residual ZZ interactions, which can impact gate fidelities in multi-qubit systems~\cite{fors2024comprehensiveexplanationzzcoupling,Gambetta,Gambetta2}. Interestingly enough, such  interaction can be suppressed when coupling qubits with opposite anharmonicity~\cite{Ku2020Suppression,Zhao2020HighContrast}.

The interplay between localized and extended correlations of qubits also provides a valuable view point for quantum simulation~\cite{cao2024}.  
Specifically, linear arrays of capacitively-coupled transmons~\cite{Koch2007Charge-insensitive} have been demonstrated to provide a quantum simulator for driven-dissipative bosonic systems with attractive interactions~\cite{Siddiqi,Ustinov}.  
In this 1D platform, the many-body localized to ergodic transition has been studied both experimentally and theoretically~\cite{Roushan,xu2018Emulating,Orell2019Probing,Pan3}, and recently the experimental capabilities have been extended to ladders~\cite{Pan} and 2D arrays~\cite{Pan2,Oliver1,Oliver2,Oliver3,Andersen2025,Shtanko2025}.
In addition, theoretical works have analyzed low-energy states of the system characterized by localized excitations of (bright) solitonic type~\cite{Mansikkamaki2021Phases,Mansikkamaki2022Beyond,solitons}.

In this work we study the localization properties in qubit arrays through  the statistical distribution of the energy level spacings of  transmons and/or capacitively shunted flux qubits (CSFQs)~\cite{Nori,Steffen,Yan2016flux}.
We employ two different approaches: the first one is based on the average level-spacing ratio, see also  Ref.~\cite{Orell2019Probing}; for the second one we consider a relative entropy of the level spacing distribution known as Kullback–Leibler divergence~\cite{Berke2022}.
Both approaches show that the repulsive (CSFQ) and attractive (transmon) cases are characterized by equivalent spectral statistics, a result that we explain by resorting to an appropriate Bose-Hubbard model.
The suppression of unwanted ZZ interactions motivates us to consider arrays with alternating values of anharmonicity, that is, mixed transmon-CSFQ devices. Interestingly, we find that the localized phase can persist up to stronger coupling strength than in arrays with uniform anharmonicity but otherwise identical parameters; the increase can be significant (over 40\% in linear arrays) for typical disorder, but becomes negligible when the disorder is stronger than the anharmonicity. We also show that increasing the mismatch between the anharmonicities beyond the value needed for optimal ZZ suppression further stabilizes the localized phase. The latter result is obtained by considering a generalized Bose-Hubbard model; hence, the investigation of such models in superconducting qubit platforms can complement their study in the context of bosonic atoms in optical lattices~\cite{Chanda2025}.

The article is organized as follows: in Sec.~\ref{sec:models}, we introduce the models investigated in this work. Then, in Sec.~\ref{sec:level-stat}, we describe the diagnostic tools we use for analyzing the level statistics. Our results on the comparison between the level statistics of uniform and alternating-sign one-dimensional qubit arrays are presented in Sec.~\ref{sec:results}. The extension of our findings to higher-connectivity arrays is briefly explored in Sec.~\ref{sec:higher-dim}. Finally, we summarize our results and discuss possible future directions in Sec.~\ref{sec:conclusions}.

\begin{figure*}
    \includegraphics[width=\linewidth]{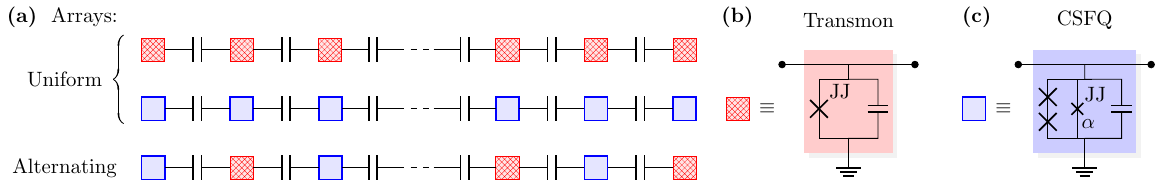}
 \caption{\textbf{(a)} The systems considered in this work: uniform arrays of transmons (top) and CSFQs (middle), and an array comprising both devices (bottom) with alternating sign anharmonicity. \textbf{(b)} Circuit diagram of a single transmon qubit. \textbf{(c)} Circuit diagram of a single CSFQ. The parameter $\alpha$ represents the ratio of the Josephson energy of the smaller Josephson junction (marked JJ) to the two identical larger Josephson junctions.}
 \label{fig:models}
\end{figure*}

\section{Models}\label{sec:models}
An array of $M$ capacitively-coupled qubits is described by the Hamiltonian 
\begin{equation}\label{eq:arrayH}
    \mathcal{\hat H} = \sum_{i=1}^M \mathcal{\hat H}^{\text{Q}}_{i} + K\sum_{i=1}^{M-1}\hat N_i\hat N_{i+1}.
\end{equation}
where $\mathcal{\hat H}^{\text{Q}}_{i}$ is the Hamiltonian of the qubit at site $i$, either a transmon ($\text{Q}=\text{T}$) or a CSFQ ($\text{Q}=\text{F}$), $K$ characterizes the coupling strength between neighboring devices, and $\hat N_i$ counts the number of excess Cooper pairs on the island of qubit $i$.
\Cref{fig:models}(a) schematically shows (top to bottom) transmon and CSFQ arrays, and an array with CSFQs/transmons at odd/even sites.

The Hamiltonian for the transmon in \cref{fig:models}(b) is~\cite{Koch2007Charge-insensitive}
\begin{equation}
\mathcal{\hat H}^{\text{T}}_{i} = 4 E_C \hat N_i^2 - E_{J i} \cos \hat{\varphi}_i,
\label{eq:transmonChainHam}
\end{equation}
where $E_C$ is the charging energy, assumed to be the same for all transmons in an array, $E_{Ji}$ is the Josephson energy, and $\hat \varphi_i$, conjugated to $\hat{N}_i$, is the phase difference across the Josephson junction at site $i$. 
We consider the transmon regime $E_{J i}\gg E_C$, in which 
the spectrum is weakly anharmonic. The site-index dependence of the Josephson energy can account for variations due to fabrication and/or design choices. We note that while $E_C$ depends on the dimensions of capacitors (typically tens to hundreds of microns) that can be fabricated accurately, the nanometer-thick tunnel barrier of the junction, determining $E_J$, is more easily subject to random fluctuations.
For the CSFQ we use the Hamiltonian~\cite{Yan2016flux} 
\begin{equation}
    \mathcal{\hat H}^{\text{F}}_{i} = 4 E_{CF} \hat N_i^2 + E_{JF i} \left\{ -2 \cos(\hat \varphi_i) + \alpha\cos(2 \hat \varphi_i) \right\},
    \label{eq:H-CSFQ}
\end{equation}
where $E_{CF}$ is the charging energy of the CSFQ, $E_{JF i}$ is the Josephson energy of the two identical large junctions of the $i^\text{th}$ CSFQ, and $\alpha$ is the ratio between the Josephson energy of the third (smaller) Josephson junction and $E_{JFi}$, see \cref{fig:models}(c). Both $E_{CF}$ and $\alpha$ are taken as independent of $i$ and we have assumed to apply an external magnetic field such that half a flux quantum threads the loop formed by the three junctions. Here $\hat{\varphi}_i$ is a collective variable, the average of the phase differences across the two large junctions, and we ignore the mode associated with the difference of the phases as it has much higher energy~\cite{Yan2016flux}.

Equation~\eqref{eq:arrayH}
can be approximately cast in the form of a generalized Bose-Hubbard model.
Specifically, we introduce bosonic annihilation, $\hat{b}_i$, and creation, $\hat{b}^\dagger_i$, operators at each site via the relations $\hat{\varphi}_i = (4 A_i)^{-1/4}(\hat b^\dagger_i + \hat b_i)$ and $\hat N_i = \imath (A_i/4)^{1/4} (\hat b^\dagger_i - \hat b_i)$~\cite{Wallraff2021Circuit}, where $A_i$ for each qubit type is given in \cref{tab:parameters}.
We then 
expand the cosines up to the fourth power of $\hat\varphi_i$ and for consistency we keep only terms that commute with the number operator $\hat{n}_i = \hat{b}^\dagger_i \hat{b}_i$ (that is, we consider terms up to the next-to-leading order in $\sqrt{1/A_i} \ll 1$).
Making a similar approximation in the term coupling neighboring qubits, where we keep the contributions that commute with the total number operator $\hat{n}=\sum_i \hat{n}_i$, we obtain
\begin{align}
\mathcal{\hat H}_\text{BH} & = \sum_{i=1}^{M} \hbar\omega_{01i} \hat{n}_i  + \sum_{i=1}^{M} \frac{U_i}{2}  \hat{n}_i (\hat{n}_i - 1) \nonumber \\ & + \sum_{i=1}^{M-1} J_{i,i+1} \left( \hat{b}_{i+1}^\dagger \hat{b}_i + \textrm{H.c.}\right) \, ,
\label{eqn:H-BH}
\end{align}
where $\hbar=h/2\pi$ is the reduced Planck constant. The parameters of the Bose-Hubbard Hamiltonian are related to those of the qubits as in \cref{tab:parameters}. Note that with our approximations the frequencies $\omega_{01i}$ and hopping coefficients $J_{i,i+1}$ depend on the site, while the interaction strengths $U_i=U^Q$ depend only on the qubit type, $U^T$ being always negative for transmons and $U^F$ for CSFQs being positive when $1/8 <\alpha < 1/2$. Note that since $[\mathcal{\hat H}_\text{BH},\hat{n}] = 0$, 
we can subtract a term $\bar{\omega}_{01} \hat{n}$ from the right hand side, where $\bar{\omega}_{01}$ is the average of the frequency over the sites; this shows that the relevant energy scales are the typical fluctuation in the frequency,
the interaction energies, and the (average) hopping amplitude. 

\begin{table}
    \centering
    \begin{tabular}{c|c|c}
        & Transmon & CSFQ \\
        \hline
        \hline
        $A_i$ & 
        $E_{J_i}/(8 E_C)$
        & 
        $(1-2\alpha)E_{JF_i}/(4 E_{CF})$
        \\
        $\hbar\omega_{01i}$ & $\sqrt{8 E_{J_i} E_C} - E_C$ & $\sqrt{16 (1-2\alpha) E_{JF_i} E_{CF}} - E_{CF}\frac{1-8\alpha}{1-2\alpha}$ \\
        $U_i$ & $-E_C$ & $E_{CF} (8\alpha-1)/(1-2\alpha)$ \\ \hline
        $J_{i,i+1}$ & \multicolumn{2}{c}{$\frac{K}{2} \sqrt[4]{A_i A_{i+1}}$} \\
    \end{tabular}
    \caption{Relations between the parameters of the qubit array Hamiltonian, Eq.~\eqref{eq:arrayH}, and the Bose-Hubbard one, Eq.~\eqref{eqn:H-BH}.}
    \label{tab:parameters}
\end{table}

\section{Diagnostic tools for level statistics}\label{sec:level-stat}
Studying the statistics of the difference between eigenenergies is a well-established tool to determine whether a system is localized or chaotic.  Poisson statistics characterizes  localized systems and Wigner-Dyson statistics chaotic ones~\cite{DAlessio2016From,MBL_RMP}. Here we consider the distribution $P(r)$ of the consecutive level spacing ratio
\begin{equation}\label{eqn:rn}
    r_n = \min\left\{\frac{E_{n+1} - E_n}{E_n - E_{n-1}},\frac{E_n - E_{n-1}}{E_{n+1} - E_n}\right\},
\end{equation}
where index $n=2,\ldots,D-1$ counts the eigenstates in ascending order of energy and $D$ is the dimension of the considered sector of the Hilbert space with a fixed number $\mathcal{N}$ of excitations. By definition, $0\le r_n \le 1$, and, more importantly for our purposes, the distribution of $r_n$ is independent of the local density of states~\cite{David2007Localization,Atas2013Distribution}. The Poisson statistics for energy levels results in the probability distribution for $r_n$ being $P_0(r) = 2/(1+r)^2$. No analytical formulas are in general available for the  Wigner-Dyson statistics.  Here, since the Hamiltonians have only real entries, we are interested in the distribution for the Gaussian Orthogonal Ensemble (GOE)~\cite{mehta2004random,guhr}, for which we use the expression
\begin{equation}
    P_{1}(r) = 2 C_1 \frac{r + r^2}{\left[(1+r)^2-0.875 r\right]^{5/2}}
    \label{eq:P-WD}
\end{equation}
where $C_1 \simeq 3.662$ is the normalization constant. The subscript $\beta=0,\,1$ in $P_\beta$ is the so-called Dyson index, denoting the absence ($\beta=0$) or presence ($\beta>0$) of level repulsion. The choice of the form of $P_1$ and its relation to the ``Wigner surmise'' are discussed in Appendix~\ref{app:fittingOfBeta} (cf. Ref.~\cite{Relano2020Distribution}).

To quantitatively characterize the crossing point between localization and chaos as we change parameters in the Hamiltonians, we consider two diagnostic tools:  \\
1. the average level spacing ratio $\bar{r} = \langle\sum_n r_n/(D-2)\rangle$, with $\langle\ldots\rangle$ denoting averaging over a Gaussian distribution of Josephson energies. The values for the Poisson and GOE distributions are $\bar{r}_0 = \int_0^1 dr\, P_0(r) r = 2 \log 2-1\simeq 0.3863$ and $\bar{r}_1 \simeq 0.5308$, and we define the crossing point by requiring that $\bar{r} = (\bar{r_0}+\bar{r}_1)/2$; \\
2. the Kullback-Leibler (KL) divergence $D_{KL}$,
\begin{equation}
D_\text{KL}(P||Q) = \sum_k p_k \log\left(\frac{p_k}{q_k}\right),
\label{eq:KLDistance}
\end{equation}
which gives the entropy of distribution $P$ relative to distribution $Q$. Here $P$ is the numerically calculated distribution $P(r)$, with index $k$ in $p_k$ denoting the $k$th bin to which the $r_n$ are assigned (see Appendix~\ref{app:simulationMethods} for more details on the numerical approach), and $Q$ is either $P_0$ or $P_1$; the crossing point is identified by requiring that $D_\text{KL}(P||P_0) = D_\text{KL}(P||P_1)$.

\begin{figure}
    \centering
    \includegraphics[width=\linewidth]{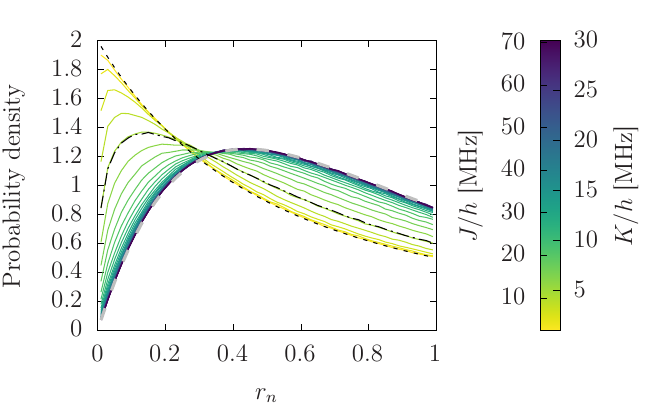}
    \caption{
    The distribution of the level spacing ratio $r_n$ [Eq.~\eqref{eqn:rn}] for a $M=10$ CFSQ array (solid lines) with $\mathcal{N}=5$ excitations for different coupling strengths $K$ in Eq.~\eqref{eq:arrayH} (color scale). The qubit parameters are $\alpha=0.35$, $E_{CF}/h=54$~MHz, and $E_{JF}$ is randomly selected from a Gaussian distribution with mean $E_{JF}/h=301$~GHz and standard deviation $\delta E_{JF}/h=8.51$~GHz; these parameters match those of transmon devices~\cite{Berke2022}. 
    The data is averaged over 5000 disorder realizations.
    The dashed lines show the distributions $P_{0}$ (black) and $P_{1}$ (gray). The dot-dashed line shows the level statistics for the Bose-Hubbard model [Eq.~\eqref{eqn:H-BH}] with average hopping amplitude $J$ equivalent to $K/h=5~$MHz. The average values of $J$ (left axis in the color bar) are computed from $K$ according to \cref{tab:parameters}.}
    \label{fig:rn}
\end{figure}

\begin{figure}
    \centering
    \includegraphics[width=\linewidth]{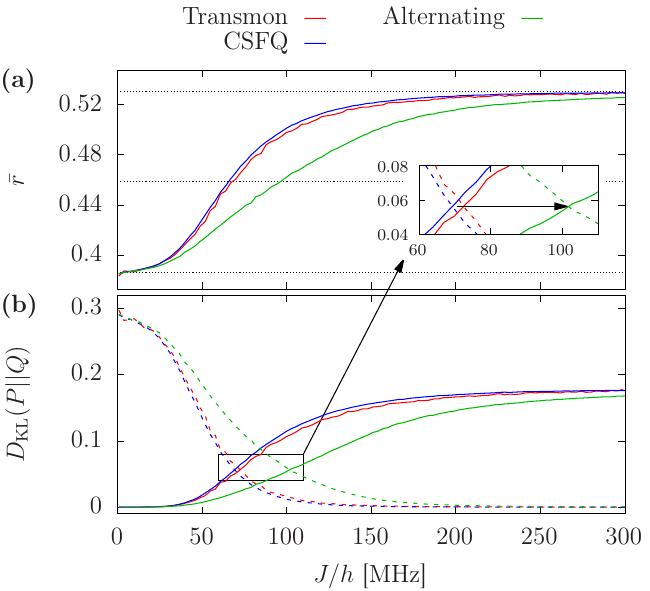}
    \caption{\textbf{(a)} Average level spacing ratio and \textbf{(b)} KL divergence 
    for transmon (red), CSFQ (blue), and alternating transmon--CSFQ (green) arrays as functions of average hopping amplitude $J$. The dotted horizontal lines in panel (a) correspond to (top to bottom) $\bar{r}_1$, $(\bar{r}_0+\bar{r}_1)/2$, and $\bar{r}_0$. In panel (b), the solid curves are for $D_\mathrm{KL}(P||P_0)$ and the dashed ones for $D_\mathrm{KL}(P||P_1)$.
    Parameters for CSFQs are as in \cref{fig:rn}, the corresponding parameters for transmons are $E_C/h=250~$MHz, $E_J/h=44~$GHz, $\delta E_J/h=1.17~$GHz~\cite{Berke2022}. 
    The inset zooms into the crossing points, highlighting the positive shift in coupling strength for the alternating array when compared to the transmon- and CSFQ-only systems.}
    \label{fig:csfq}
\end{figure}

\section{Level statistics in one-dimensional arrays}\label{sec:results}
Having introduced the relevant models and the diagnostic tools in Secs.~\ref{sec:models} and \ref{sec:level-stat} respectively, we move to the discussion of the level statistics for some concrete cases. To minimize finite-size effects, we focus on one-dimensional arrays in this section, while extensions to two-dimensional arrays are considered below in Sec.~\ref{sec:higher-dim}. 

\Cref{fig:rn} shows the disorder-averaged distribution of $r_n$ for a CSFQ array [Eq.~\eqref{eq:arrayH} with $Q=F$] for a range of coupling strengths. We consider the case of half filling (the total number of excitations being half the length of the array), as states with maximum local occupancy 1 within this subspace are expected to be representative of the computational subspace~\cite{Berke2022}. The distribution nearly follows $P_{0}$ (localized) for the lowest coupling considered, $K/h=1~$MHz, and $P_{1}$ (chaotic) for the largest, $K/h=30~$MHz. With increasing coupling strength, the distributions evolve from $P_{0}$ to $P_{1}$. This evolution can be investigated using so-called intermediate statistics~\cite{Intermediate}; here we only point out that,
similarly to the case of transmon arrays~\cite{Orell2019Probing}, we find that the level statistics of the Bose-Hubbard model [Eq.~\eqref{eqn:H-BH}] agrees with that of the CSFQ array, as we show explicitly for one intermediate coupling value. 

Now turning to our diagnostic tools, we plot in \cref{fig:csfq} the average level spacing and KL divergences for transmon [red curve , $Q=T$ in Eq.~\eqref{eq:arrayH}], CSFQ (blue curve, $Q=F$), and alternating transmon--CSFQ (green curve, $Q=T/F$ for even/odd site) arrays as a function of the qubit-qubit coupling strength expressed in terms of the (average) hopping amplitude $J$. The qubit parameters are chosen such that the average frequency $\bar{\omega}_{01}$ as well as its standard deviation $\delta\omega_{01}$ are the same for all three array types, and the anharmonicities are opposite for transmons and CSFQs, $U^F=-U^T\equiv U$. Both $\bar{r}$ and $D_\mathrm{KL}$ change smoothly with coupling strength (up to fluctuations whose amplitude decreases by increasing the number of disorder realizations). Clearly, at low coupling all of the distributions are in the localized regime, since $\bar{r} \simeq r_0$ and $D_\mathrm{KL}(P||P_0) \ll D_\mathrm{KL}(P||P_1)$, and at large coupling in the chaotic one, $\bar{r} \simeq r_1$ and $D_\mathrm{KL}(P||P_0) \gg D_\mathrm{KL}(P||P_1)$.

The transmon and CSFQ arrays display remarkably similar dependencies of $\bar{r}$ and $D_\mathrm{KL}$ on $J$ (the transmons' $D_\mathrm{KL}$ curves would agree with those in Ref.~\cite{Berke2022} if plotted with the same normalization); both quantities are found to be close to the values obtained for the Bose-Hubbard model. This feature reflects a remarkable symmetry that holds for the level statistics of Bose-Hubbard model under the exchange $U\leftrightarrow -U$,
if the same disorder distribution symmetric around $\bar{\omega}_{01}$ is used for the two signs of $U$ (the derivation is given in Appendix~\ref{app:dBHmLevelStats}). 
For all three qubit array types, the crossing points in panel (a) of \cref{fig:csfq} (from $\bar{r})$ are comparable to those in panel (b) (from $D_\mathrm{KL}$). This finding cross-validates the two approaches. For concreteness, in what follows we define the crossing point hopping amplitude $J_C^Q$ as the value of $J$ for which the divergences are the same for arrays of type $Q=T,\,F,\,A$ (with $A$ denoting arrays with alternating transmon/CSFQ qubits). 

\begin{figure}
    \centering
    \includegraphics[width=\linewidth]{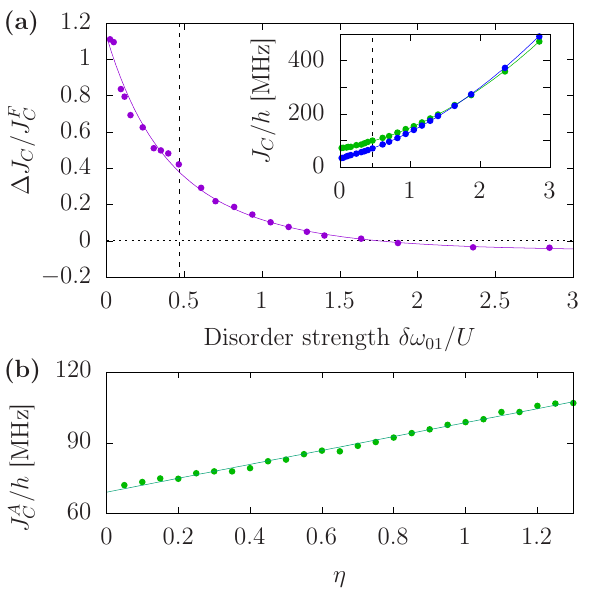}
    \caption{\textbf{(a)} Relative shift in crossing point $\Delta J_C / J_C^F$ as a function of the disorder strength $\delta \w_{01}/U$ for the Bose-Hubbard model.  The dashed vertical line shows the disorder strength used in all other figures (cf. Ref.~\cite{Berke2022}).
    The inset shows the numerically calculated values of $J_C^F$ (blue symbols) and $J_C^A$ (green symbols), fitted by quadratic functions (solid curves);
    the relative difference between those two curves gives the solid line in the main panel.
    \textbf{(b)} Crossing point hopping amplitude $J_C^A$ for an array with alternating transmons and CSFQs vs. ratio of anharmonicites $\eta = -U^F/U^T$ (we change $U_F$ while keeping fixed $U_T=-250~{\rm MHz}\ h$, other parameters as in \cref{fig:csfq}, mapped to the Bose-Hubbard model according to Table~\ref{tab:parameters}). The linear best-fit line is shown as a guide to the eye.
    The points in all panels are obtained by averaging over 1000 disorder realizations.}
    \label{fig:disorder-scaling-alt}
\end{figure}

The inset in \cref{fig:csfq}(a) highlights one of our main results, namely that, for disorder strength as in typical devices, the crossing point hopping amplitude is larger for alternating arrays, $J^A_C > J_C^T \simeq J_C^F$. 
We quantify this enhancement by introducing $\Delta J_C = J^A_C - J^F_C$; in \cref{fig:disorder-scaling-alt}(a) we plot $\Delta J_C/J_C^F$ [calculated for the Bose-Hubbard model, Eq.~\eqref{eqn:H-BH}] as function of $\delta\omega_{01}/U$. The relative enhancement of $J_C$ is largest for weak disorder and decreases monotonically with disorder strength. Such a result can be qualitatively understood by considering the level repulsion between next-nearest-neighbor sites: at small disorder, the fact that the multi-excitation levels of the site placed between the next-nearest-neighbors are detuned due to the opposite sign of $U$ favors localization, but this effect becomes less important with increasing disorder (see Appendix~\ref{app:3sites}).
In fact, for strong disorder, $\delta\omega_{01}/U \gtrsim 1.7$, we find that $J_C^A< J_C^F$, even though for all array types the crossing point hopping increases with disorder, see the inset in \cref{fig:disorder-scaling-alt}(a).

So far, we have set $U^T = -U^F$ for the alternating arrays, a choice motivated by the cancellation of ZZ interactions between neighboring qubits~\cite{Zhao2020HighContrast,Ku2020Suppression}. However, from the point of view of localization, we find this is not necessarily the optimal choice. This remarkable feature is highlighted  in \cref{fig:disorder-scaling-alt}(b). Indeed, $J_C$ in an alternating array can be increased or decreased by changing the ratio $\eta = -U^F/U^T$. In other words, our results show that the localization enhancement in arrays with alternating sign anharmonicity is unrelated to
the suppression of the residual ZZ interaction in such systems.

\section{Level statistics of two-dimensional arrays}\label{sec:higher-dim}

\begin{figure}
    \centering
    \includegraphics[width=\linewidth]{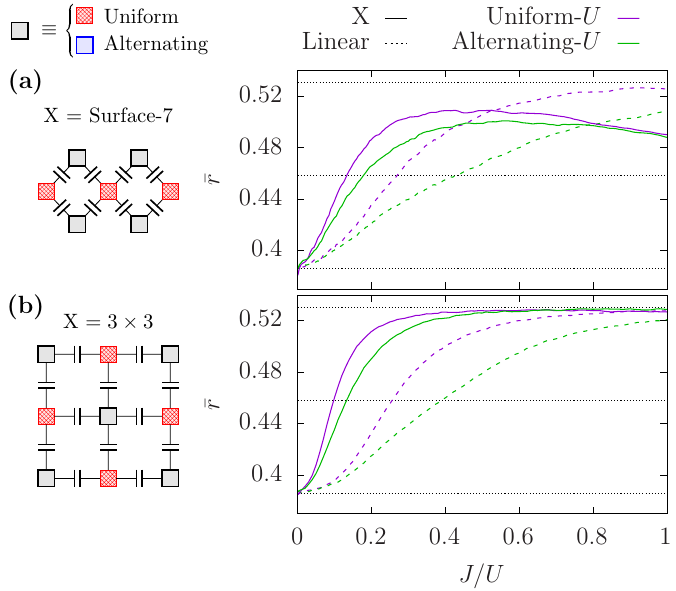}
    \caption{
    Level statistics in two-dimensional arrays. The solid curves show the average level spacing ratio $\bar{r}$ as function of the hopping amplitude $J$ for \textbf{(a)} a surface-7 chip with $\mathcal{N}=4$ particles and \textbf{(b)} a $3\times3$ lattice with $\mathcal{N}=5$ particles for uniform (purple) and alternating-sign interaction (green). For comparison, the average level spacing ratio for linear chains with the same particle and site numbers are displayed with dashed lines. The dotted horizontal lines in both panels correspond to (top to bottom) $\bar{r}_1$, $(\bar{r}_0 + \bar{r}_1)/2$, and $\bar{r}_0$. For each value of $J$ (assumed homogeneous throughout the array), we average over $1000$ disorder realizations taken from a Gaussian distribution with standard deviation $\delta\omega/|U|=0.47$ for the Hamiltonian in Eq.~\eqref{eqn:H-BH}, where additional hopping terms are included according to the connectivity schematized on the left of the panels.
    }
    \label{fig:dimensions}
\end{figure}

In the previous section we showed that one-dimensional alternating-sign anharmonicity qubit architectures are more resilient to chaos; the arrays used in actual devices for quantum computing purposes are usually two-dimensional~\cite{Wallraff2021Circuit}, so here we explore to what degree higher connectivity affects our results. For the computation of the full spectrum we use the exact diagonalization technique, which is typically limited to small size and few particles, $M\sim 10-12$ at half-filling~\cite{Orell2019Probing,DiVincenzo3} since $D_{M/2}^M\propto 2.6^M/\sqrt{M}$ for large $M$, and so not particularly suitable for higher-dimensionality arrays (here $D_{\mathcal{N}}^M=\binom{\mathcal{N}+M-1}{\mathcal{N}}$ is the dimension of the $\mathcal{N}$-particle sector). However, we can explore two minimal yet relevant case studies, in analogy with the discussion given in~\cite{Berke2022}, i) the surface-7 chip architecture~\cite{Andersen2020}, and ii) a $3\times 3$ array. The qubit connections for the two cases are shown on the left of Fig.~\ref{fig:dimensions}(a) and (b), respectively; for instance, the surface-7 chip is obtained from a linear chain of seven qubits by connecting the middle one (qubit 4) to the first and the last ones (qubits 1 and 7), and so the model Hamiltonian is obtained by adding the terms $J_{4,1} \hat{b}_4^\dagger b_1+J_{4,7}\hat{b}_4^\dagger b_7+{\rm H.c.}$ to the Hamiltonian in Eq.~\eqref{eqn:H-BH}. Our goal is to compare the level statistics in uniform and alternating-sign anharmonicity arrays, the latter case having qubits with opposite anharmonicity ($\eta=1$) as shown in Fig.~\ref{fig:dimensions}.

The solid curves in the main panels of Fig.~\ref{fig:dimensions} show the average level spacing ratio $\bar{r}$ as function of the hopping $J$ (in units of $U$) for the surface-7 chip with $\mathcal{N}=4$ (panel a) and a $3\times 3$ array with $\mathcal{N}=5$ (panel b), both for a transmon array (purple) and an alternating transmon–CSFQ array (green). For comparison, we also display the average level spacing ratio for linear chains with the same $\mathcal{N}$ and $M$ (dashed curves in Fig.~\ref{fig:dimensions}). The crossing points for the surface-7 chip and the $3\times3$ array -- the intercepts between the curves and the middle horizontal line, see main text -- are sizably reduced compared to the linear case, since the higher connectivity suppresses the localization effects (analog results have been reported in Ref.~\cite{Berke2022} for transmon-only arrays).  Nonetheless, the use of alternating transmon-CSFQ arrays still leads to stronger resilience to chaos, although in a reduced fashion: for the surface-7 chip the relative increase of the crossing point, $(J_C^A-J_C^F)/J_C^F$, decreases from $59\%$ to $34\%$, and for the $3\times3$ array from $52\%$ to $36\%$. These results confirm that the use of alternating-sign anharmonicity qubits is beneficial also for two-dimensional arrays. 

\section{Conclusions}\label{sec:conclusions}
We have investigated the crossover from localized to chaotic regimes in arrays of superconducting qubits comprising C-shunted flux qubits (CSFQ) and transmon qubits displaying positive and negative anharmonicity, respectively, as described by the Hamiltonians in Eqs.~\eqref{eq:arrayH}-\eqref{eq:H-CSFQ}. Once the parameters of the models are appropriately  matched, the statistics of the level spacing ratio can also be reproduced by using the disordered Bose-Hubbard model of Eq.~\eqref{eqn:H-BH}. We note that the level statistics of the Bose-Hubbard model is found to be independent of the sign of the onsite interaction $U$. Such finding is in contrast with the ground-state properties of the model being clearly different in the attractive or repulsive  cases (see Ref.~\cite{Mansikkamaki2021Phases} and references therein).

Our results indicate that for small disorder strength relative to the anharmonicity, arrays with alternating transmons and CSFQs  remain localized up to a higher coupling strength compared with chains consisting of one qubit type only, see Figs.~\ref{fig:csfq} and \ref{fig:disorder-scaling-alt}. Interestingly, the onset of the chaotic behavior arises at even higher coupling when increasing the difference between nearest-neighbor anharmonicities. For large disorder strength compared to the anharmonicity, arrays with a single qubit type or alternating qubits display similar localization properties.

For quantum computing applications, future research could extend our investigation of the localized to chaotic crossover to additional diagnostic tools, such as the Walsh transform~\cite{Berke2022}. Larger systems, especially in higher dimensions, could be studied via a classical approach~\cite{DiVincenzo3}. On the quantum simulation side, the possibility of engineering the strength and nature (attractive/repulsive) of the on-site interaction provides a new platform for the investigation of many-body localization. More broadly,
we highlight that superconducting qubits enable tailoring the interaction properties of the system at a {\it local} level, beyond what has been achieved so far in other platforms, for example the cold atoms one~\cite{gross2017quantum,schafer2020tools}.

\section*{Acknowledgments}
Specialist and High Performance Computing systems provided by Information Services at the University of Kent.

\appendix

\section{Distributions for the consecutive level spacing
ratio}\label{app:fittingOfBeta}

The study of the distribution of level spacings in quantum systems has a long history, in particular within Random Matrix Theory~\cite{mehta2004random,guhr}.  The well-known generalization of 
the Wigner surmise for the level spacing distribution $P_W(s)$ of random matrices, corresponding to exact results for $2\times 2$ matrices, is
\begin{equation}
    P_W(s) = a_\beta s^\beta e^{-b_\beta s^2},
\end{equation}
where $s_n = E_{n+1}-E_n$, and $a_\beta$ and $b_\beta$ are known constants~\cite{guhr}. The Dyson index $\beta$ depends on the symmetries of the considered random matrices, with $\beta=1$ for the Gaussian Orthogonal Ensemble (GOE) of real symmetric matrices, $\beta=2$ for the Gaussian Unitary Ensemble (Hermitian matrices), and $\beta=4$ for the Gaussian Symplectic Ensemble (Hermitian quaternionic matrices).

\begin{figure}
    \centering
    \includegraphics[width=\linewidth]{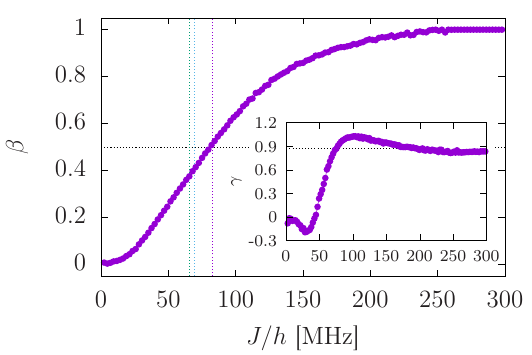}
    \caption{The values of $\beta$ (main panel) and $\gamma$ (inset) found by fitting \cref{eqn:Pgammabeta} to numerical level spacing ratio distributions for the uniform-interaction Bose-Hubbard as a function of coupling strength $J$. The dashed horizontal line is at $\beta=0.5$, and the intersection with the data gives the value of the crossing point between localized and chaotic regimes $J_C$ (dashed purple vertical line). For comparison, we also show the crossing points determined with the methods of the main text, the $D_\text{KL}$ crossing (dashed green line) and $\bar{r}=0.4585$ (dashed blue line). The horizontal line in the inset is at $\gamma=0.875$.
    Parameters are as in \cref{fig:rn}, mapped to the Bose-Hubbard model according to \cref{tab:parameters}, with 5000 disorder realizations used.
    }
    \label{fig:bhm-beta}
\end{figure}

\begin{figure}
    \centering
    \includegraphics[width=\linewidth]{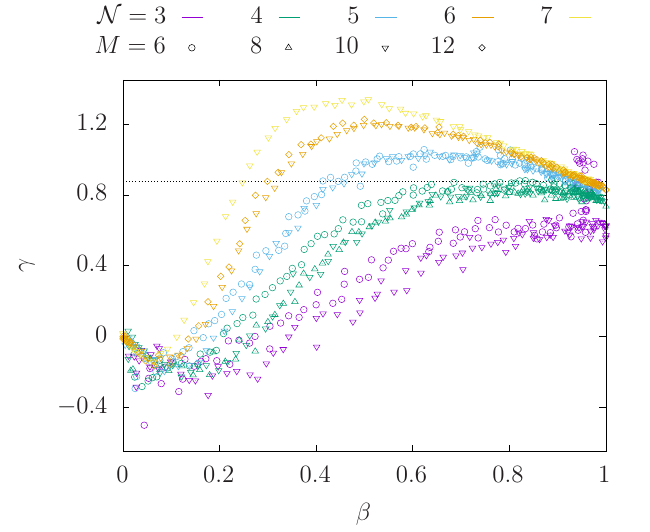}
    \caption{ 
    Plot of $\gamma$ as a function of $\beta$, both found by two-parameter fits of \cref{eqn:Pgammabeta} to numerical level spacing ratio distributions for the uniform-interaction Bose-Hubbard model [\cref{eqn:H-BH}], for various numbers of excitations $\mathcal{N}$ and sites $M$. Each point corresponds to one value of $J$, with 1000 disorder realizations used.
    Other parameters are as in \cref{fig:rn} and mapped to the Bose-Hubbard model according to \cref{tab:parameters}. The dashed horizontal line shows $\gamma=0.875$.
    }
    \label{fig:r_n_beta_AND_gamma_beta}
\end{figure}

In analogy with the Wigner surmise, an approximate distribution $P_W(r)$ for the level spacing ratios [\cref{eqn:rn}], which corresponds to the exact distribution for $3\times 3$ random matrices, was derived in Ref.~\cite{Atas2013Distribution},
\begin{equation}
    P_W(r) = \frac{2}{Z_\beta} \frac{(r+r^2)^\beta}{(1+r+r^2)^{1+3\beta/2}},
    \label{eqn:pw}
\end{equation}
where $Z_1=8/27$ for the GOE. 
To be able to both capture the distribution for large matrices and interpolate between localized and chaotic regimes, a generalization of \cref{eqn:pw} was proposed in Ref.~\cite{Relano2020Distribution},
\begin{equation}
    P_\beta(r;\gamma) = 2 C_{\beta\gamma} \frac{(r + r^2)^\beta}{\left[(1+r)^2-\gamma r\right]^{1+3\beta/2}},
    \label{eqn:Pgammabeta}
\end{equation} 
where $C_{\beta\gamma}$ is a normalization coefficient such that $\int_0^1 dr\, P_\beta(r;\gamma)=1$. \Cref{eqn:Pgammabeta} with $\beta=0, \gamma=0$ reduces to the distribution $P_0$ of the main text, and with $\gamma=1$ to $P_W(r)$ in \cref{eqn:pw}.
Since the models which we study in this work only have real entries in the Hamiltonian, we are interested in the GOE, $\beta=1$; to find corresponding value for $\gamma$, we fit $P_1(r;\gamma)$ to estimates for the distribution obtained from numerical data for large matrices in Refs.~\cite{Atas2013Distribution} and \cite{Relano2020Distribution}, finding $\gamma \simeq 0.875$ (cf. \cref{fig:bhm-beta,fig:r_n_beta_AND_gamma_beta}), see \cref{eq:P-WD} of the main text.

As mentioned above, \cref{eqn:Pgammabeta} can interpolate between the localized and chaotic regimes~\cite{Relano2020Distribution}; here we use this feature to capture the localized--chaotic crossing point by assuming $\beta \in [0\ldots 1]$. We perform a two-parameter fit of \cref{eqn:Pgammabeta} to the binned frequency distributions at each value of hopping amplitude $J$, such as those shown in \cref{fig:rn}, to find $\beta$ and $\gamma$. \Cref{fig:bhm-beta} shows the value of the fitted $\beta$ as a function of $J$ for the Bose-Hubbard (BH) model in \cref{eqn:H-BH} with uniform interaction. The index $\beta$ monotonically increases with $J$ from $0$ (localized regime) to $1$ (chaotic); we define the crossing point hopping $J_C$ as the value of $J$ for which $\beta=0.5$, identified by the purple dashed vertical line. The other vertical dashed lines in \cref{fig:bhm-beta} show the crossing points estimated using the methods used in the main text, the $D_\text{KL}$ crossing (green) and $\bar r$ (blue): the value of $J_C$ from $\beta=0.5$ is approximately $13\,$MHz higher than that from the $D_\text{KL}$ crossing, which in turn is about $3\,$MHz higher than that from the $\bar{r}$ method.

\begin{figure}
    \centering
    \includegraphics[width=\linewidth]{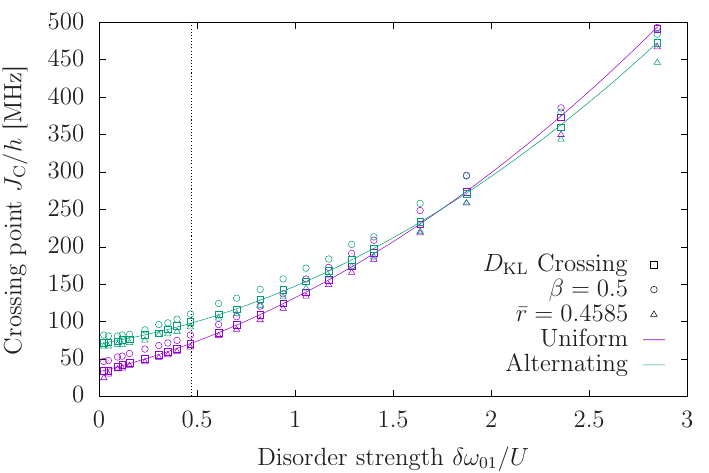}
    \caption{
    The effect of disorder strength on the crossing point $J_C$, extracted from the $D_\text{KL}$ crossing (squares), $\beta=0.5$ (circles), and $\bar r=0.4585$ (triangles), for the uniform-interaction (purple) and alternating-interaction (green) Bose-Hubbard model. 
    The solid lines are quadratic polynomial fits to the $D_\text{KL}$ crossing values.
    The dashed vertical line represents the disorder strength used throughout this work (cf. \cref{fig:disorder-scaling-alt}). Parameters are as in \cref{fig:rn} in the main text. Data is for 1000 disorder realizations.
    }
    \label{fig:crossover-comparison}
\end{figure}

Despite the quantitative differences in the estimates for $J_C$, all three methods give a consistent picture of the evolution of the crossing point with disorder, as shown in \cref{fig:crossover-comparison}: for both uniform and alternating interaction in the BH model, $J_C$ increases monotonically with disorder strength $\delta\omega_{01}$. Moreover, all three methods agree on the value of $\delta\omega_{01}/U\sim 1.7$ above which $J_C$ for the uniform case become larger than for the alternating one.

In closing this section, we briefly discuss a potential limitation in using \cref{eqn:Pgammabeta} to study the localized to chaotic crossover. 
In proposing Eq.~\eqref{eqn:Pgammabeta}, the authors of Ref.~\cite{Relano2020Distribution}
already noted that the relation between $\gamma$ and $\beta$ is not universal, since it depends on the model being studied: 
if there were a unique relationship determining $\gamma$ as function of $\beta$, then the crossover could be analyzed in terms of $\beta$ only. 
Our results in \cref{fig:r_n_beta_AND_gamma_beta} show not only that for the BH model the relation differs from that of other models investigated in Ref.~\cite{Relano2020Distribution}, but also that it displays a marked dependence on the number of excitations {$\mathcal{N}$} but not on system size $M$.

\section{Numerical methods}\label{app:simulationMethods}

For the numerical studies of arrays with uniform and mixed qubit types, we consider a fixed number of excitations $\mathcal N$. We first diagonalize the Hamiltonian $\mathcal{\hat H}^{\text{Q}}_{i}$ of each single qubit at site $i$, where the $\mathcal N$ lowest eigenstates $|u\rangle_i$, $u=0,\ldots, \mathcal{N}$, with energy $E^{Qi}_{u}$ are computed in the charge basis (eigenstates of operator $\hat N_i$) restricted to $ N_i\in \{-50\ldots50\}$.
We then consider an array of qubits, \cref{eq:arrayH}, for which the Hilbert space is the tensor product of the individual qubits' Hilbert spaces, and keep the leading-order particle-number-conserving hopping terms in the expansion of $\hat{N}_i \hat{N}_{i+1}$, resulting in the Hamiltonian
\begin{align}
    \hat{\mathcal{H}}_\text{C} = &\sum_{i=1}^{M} \sum_{u=0}^{\mathcal N} E^{Qi}_{u} |u\rangle_{i}\langle u|_{i} + \nonumber\\
    &K \sum_{i=1}^{M-1} \sum_{u=0}^{\mathcal{N}-1}\sum_{v=1}^{\mathcal{N}} g_{iuv} |u+ 1\rangle_{i} \langle u|_{i} \otimes \nonumber\\
    &\hspace{40pt}|v- 1\rangle_{i+1}\langle v|_{i+1} + \text{H.c.}
    \label{eqn:H-TC-RWA}
\end{align}
with $g_{iuv} = \langle u+ 1|_{i} \hat N_i|u\rangle_{i}\langle v-1|_{i+1} \hat N_{i+1}|v\rangle_{i+1}$.

The values of $r_n$ are computed from the eigenvalues found by exact diagonalization of \cref{eqn:H-TC-RWA} and are averaged over many realizations of disorder in Josephson energy, each randomly selected from a Gaussian distribution with means $E_J$ or $E_{JF}$ and a standard deviation computed such that the distributions of $\w_{01}$ for the two devices match (see below). To calculate the KL divergence, \cref{eq:KLDistance}, we bin the resulting data into 50 equally wide bins, and the values of $p_k$ are the frequencies in each bin, while $q_k=P_i(r_k)$, $i=0,\,1$, is computed with $r_k$ being the midpoint of the $k^\text{th}$ bin.

As indicated above, to make a suitable comparison between transmon and CSFQ arrays, or to attain the cancellation of residual ZZ interaction, some care must be taken in selecting the parameters of the qubits. For transmons, we simply used the same (experimentally motivated) parameters as in Ref.~\cite{Berke2022}. For CSFQs, we need to determine values of $E_{JF}$, its standard deviation $\delta E_{JF}$, $E_{CF}$, and $\alpha$ such that the average frequency $\bar{\omega}_{01}$, its standard deviation $\delta\omega_{01}$, and (the absolute value of) the anaharmonicity $\mathcal{A}$ match those of the transmons; the anharmonicity of a qubit is given by $\mathcal{A}=\hbar(\w_{12}-\w_{01})$, where $\hbar\w_{ij}=E^Q_{j}-E^Q_{i}$.

To numerically find suitable parameter values, let us consider the following approximate expressions for $\omega_{01}$ and $\mathcal{A}$:
\begin{widetext}
\begin{align}
\label{eq:w01Approx}
\hbar\w_{01}&\approx \sqrt{16 E_{JF}(1-2\alpha) E_{CF}}-E_{CF} \frac{1-8\alpha}{1-2\alpha}+E_{CF}\sqrt{\frac{4 E_{CF}}{E_{JF}(1-2\alpha)}}\left[\frac{1}{8}\frac{1-32\alpha}{1-2\alpha}-\frac{1}{4}\left(
\frac{1-8\alpha}{1-2\alpha} 
\right)^2\right],\\
\label{eq:anarApprox}
\mathcal{A}&\approx -E_{CF} \frac{1-8\alpha}{1-2\alpha}+E_{CF}\sqrt{\frac{4 E_{CF}}{E_{JF}(1-2\alpha)}}\left[\frac{1}{4}\frac{1-32\alpha}{1-2\alpha}-\frac{17}{32}\left(
\frac{1-8\alpha}{1-2\alpha} 
\right)^2\right],
\end{align}
\end{widetext} 
valid for $E_{JF} (1-2\alpha) \gg E_{CF}$. Equations~\eqref{eq:w01Approx} and \eqref{eq:anarApprox} can be also used for the transmon with the replacements: $\alpha\to0$, $E_{CF}\to E_C$ and $E_{JF}\to E_J/2$ [compare Eqs.~\eqref{eq:transmonChainHam} and \eqref{eq:H-CSFQ}]. To determine starting points for our numerical determination of the CSFQ's parameters, we drop the last term in each of the two equations and invert them to find
\begin{align}
    E_{JF} &\approx \frac{8 \alpha - 1}{1 - 2 \alpha}\frac{\left(\sqrt{8 E_C E_J} - 2 E_C \right)^2}{16 E_C (1 - 2\alpha)}; \label{eqn:ejfcsfq}\\
   E_{CF} &\approx  E_C \frac{1 - 2 \alpha}{8 \alpha - 1}.\label{eqn:ecfcsfq}
\end{align}
Note that Eq.~\eqref{eqn:ecfcsfq} can be satisfied only for $1/8<\alpha<1/2$. Similarly, for the standard deviation $\delta E_{JF}$ using \cref{eqn:ejfcsfq,eqn:ecfcsfq} we obtain the following expression for its starting value:
\begin{equation}
\frac{\delta E_{JF}}{E_{JF}} \approx \frac{2\ \hbar\delta \w_{01}}{\hbar\w_{01} -E_C}.
\end{equation}
Setting $\alpha=0.35$ roughly in the middle of the allowed range, we  find experimentally realistic values for $E_{JF}$, cf. Ref.~\cite{Yan2016flux,Chang2022,Chang2023}.

\section{Level statistics of the Bose-Hubbard model}\label{app:dBHmLevelStats}

The Hamiltonian for the disordered Bose-Hubbard model [cf. Eq.~\eqref{eqn:H-BH}] can be written in the form
\begin{align}
\mathcal{\hat H}_\text{BH}& = \sum_{i=1}^{M} \hbar\delta\omega_{01i} \hat{n}_i  + \frac{U}{2} \sum_{i=1}^{M}  \hat{n}_i (\hat{n}_i - 1) \nonumber \\ & +  \sum_{i=1}^{M-1} J_{i,i+1}\left( \hat{b}_{i+1}^\dagger \hat{b}_i + \textrm{H.c.}\right) . 
\end{align}
where $\delta\omega_{01i}$ is the deviation of the frequency at site $i$ from the average frequency of the array (subtracting the average frequency, $\mathcal{\hat H}_\text{BH} \rightarrow \mathcal{\hat H}_\text{BH} - \bar{\omega}_{01} \hat{n}$, does not change the level spacings when considering a sector of the Hilbert space with a given number of excitations). 
Clearly, $\mathcal{\hat H}_\text{BH}' = - \mathcal{\hat H}_\text{BH}$ has the same level spacings as $\mathcal{\hat H}_\text{BH}$, the opposite frequencies, and the opposite sign for the interaction term; the sign of $J_{i,i+1}$ is unimportant and can be changed by redefining $\hat{b}_i \rightarrow (-1)^i \hat{b}_i$ (note that this transformation preserves both $\hat{n}_i$ and the bosonic commutation relations). Thus, differences in the level spacing ratios under the exchange $U\leftrightarrow -U$ are only due to the frequencies. For a distribution of frequencies symmetric around the mean, for each set of $\delta\omega_{01i}$ there exists a corresponding set with opposite values of $\delta\omega_{01i}$; in averaging over disorder (that is, over frequencies), both sets contribute to the level spacing statistics, and hence the cases with opposite values of $U$ have the same statistics, as stated in the main text.

\begin{figure}[t]
    \centering
    \includegraphics[width=\linewidth]{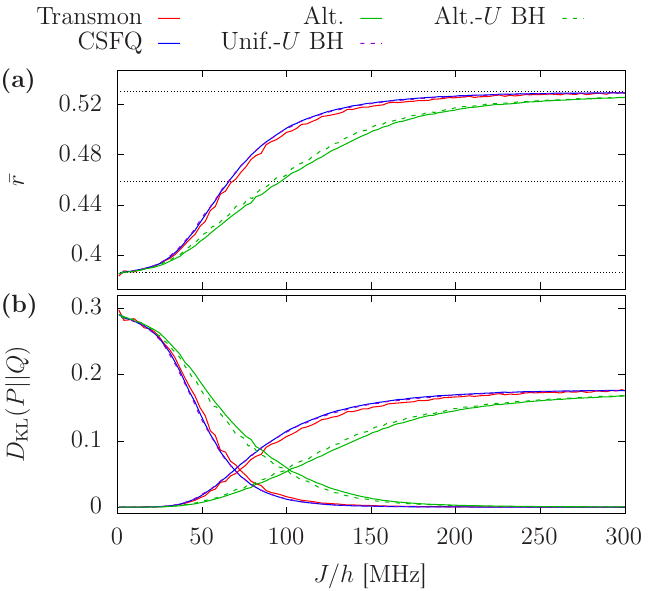}
    \caption{Crossover in qubit arrays vs Bose-Hubbard model.
    \textbf{(a)} Average level spacing ratio and \textbf{(b)} KL divergence 
    for transmon (solid red), CSFQ (solid blue), alternating transmon--CSFQ (solid green) arrays,  uniform interaction (dashed purple) and alternating-sign interaction (dashed green) BH model as functions of (average) hopping amplitude $J$. The dotted horizontal lines in panel (a) correspond to (top to bottom) $\bar{r}_1$, $(\bar{r}_0+\bar{r}_1)/2$, and $\bar{r}_0$. In panel (b), the increasing curves are for $D_\mathrm{KL}(P||P_0)$ and the decreasing ones for $D_\mathrm{KL}(P||P_1)$.
    The parameters used are as in \cref{tab:parameters} and mapped for the BH model as in \cref{fig:csfq}, with 5000 disorder realizations used. 
    }
    \label{fig:dBHm}
\end{figure}

In \cref{fig:dBHm}, we show the same data as in \cref{fig:csfq} (solid lines) and, for comparison, the corresponding results for the BH model (dashed), both with uniform and alternating interaction. 
We recall that the BH model is expected to be a good approximation to transmon and CSFQ arrays in the limit $E_J\gg E_C$, (for the CSFQ, replace $E_J\to 2(1-2\alpha)E_{JF},E_C\to E_{CF}$, cf.~\cref{tab:parameters}); indeed, we find that all quantities are within a few percent of each other. In fact, the uniform-interaction BH model curves are hardly distinguishable from those for the CSFQ array, whose $E_J/E_C$ ratio is larger than for the transmon array.

\subsection{Counter-rotating terms}\label{app:counter-rotating}

The results presented in this work are obtained using the rotating-wave approximation~\cite{Wallraff2021Circuit}; in other words, terms which do not preserve the total number of excitations are neglected [cf. Eq.~\eqref{eqn:H-BH}]. This approximation holds for small coupling compared to the average qubit frequency $J/\bar{\omega}_{01}\ll 1$, with relative corrections to the energy levels being $\sim (J/\bar{\omega}_{01})^2$ at the leading order in perturbation theory. Here we show that the counter-rotating (CR) terms give a negligible contribution to the level statistics of the investigated models. For concreteness, we focus on the disordered Bose-Hubbard Hamiltonian, and include the CR terms as follows: 
\begin{equation}\label{eq:bhm-cr}
    \mathcal{\hat H}_\text{BH-CR} = \mathcal{\hat H}_\text{BH} + \sum_{i=1}^{M-1} J_{i, i+1} \left( \hat{b}_{i+1}^\dagger \hat{b}^\dagger_i 
    + \hat{b}_{i+1} \hat{b}_{i}\right) \,,
\end{equation}
where the first (second) term in the round parentheses raises (lowers) excitation numbers of two neighboring sites by one excitation each. Thus, last term of the Hamiltonian in Eq.~\eqref{eq:bhm-cr} does not conserve the total number of excitations, $[\hat{\mathcal{H}}_{\rm BH-CR},\hat{n}]\neq 0$, and connects all the sectors of the Hilbert space with the same parity, yielding an infinite-dimensional Hilbert space which needs to be truncated in numerical calculations. We account for the effect of the CR terms on the spectrum of the $\mathcal{N}$-particle sector [with dimension $D_{\mathcal N}^M=\binom{\mathcal{N}+M-1}{\mathcal{N}}$] at the second-order in perturbation theory. Concretely, we consider the matrix representation of $\hat{\mathcal{H}}_{\rm BH-CR}$ on a basis set including the $\mathcal{N}$-particle sector Fock states $|\psi_k\rangle$ with $k=1,\ldots,D_{\mathcal N}^M$, and all the states $|\phi_l\rangle$ with $l=1,\ldots,\tilde{D}_{\mathcal N}^M$ for which $\langle \phi_k|\sum_{i=1}^{M-1} \left( \hat{b}_{i+1}^\dagger \hat{b}^\dagger_i + \hat{b}_{i+1} \hat{b}_{i}\right)|\psi_l\rangle\neq 0$ for at least one $k$. Clearly, these states belong to the $\mathcal{N}-2$ or the $\mathcal{N}+2$-particle sectors; their number ($\tilde{D}_{\mathcal N}^M$) can be evaluated in a closed form, reading (for $M> 2$ and $\mathcal{N}\geq 2$)
\begin{equation}
\tilde{D}_{\mathcal N}^M=(M-1) D_{\mathcal N}^M+
\frac{(\mathcal{N}+M-4)(\mathcal{N}+M-3)}{M-2}D_{\mathcal N-2}^{M-2}
\label{eq:additionalStates}
\end{equation}
and $\tilde{D}_{\mathcal N}^{M=2}=2\mathcal N$ for $M=2$. The first (second) term on the right-hand side of Eq.~\eqref{eq:additionalStates} accounts for the states reached by acting on the $\mathcal{N}$-particle sector with the two creation (annhiliation) operators in the CR term. This formula shows that the dimension of the Hilbert space we need to consider to include perturbatively the CR term is at least $M$ times that of the original one, and so the computation of the spectrum is more costly. Taking for example the case $\mathcal{N}=4$, $M=10$, we have $D_{\mathcal N}^M=715$ and $\tilde{D}_{\mathcal N}^{M}=6930$, and the dimension of the Hilbert space increases by approximately one order of magnitude. After computing the $D_{\mathcal N}^{M}+\tilde{D}_{\mathcal N}^{M}$ eigenvalues for each disorder realization, the level statistics is computed by selecting the $D_{\mathcal N}^{M}$ eigenvalues for which the expectation value of the total particle number operator $\hat n$ on the corresponding eigenstates rounds to $\mathcal{N}$; this procedure  is robust because the relative deviation $\left(\langle \hat n\rangle-\mathcal{N}\right)/\mathcal{N}$ over the these states is of order $J^2/\bar\omega_{01}^2$.

\begin{figure}
    \centering
    \includegraphics[width=\linewidth]{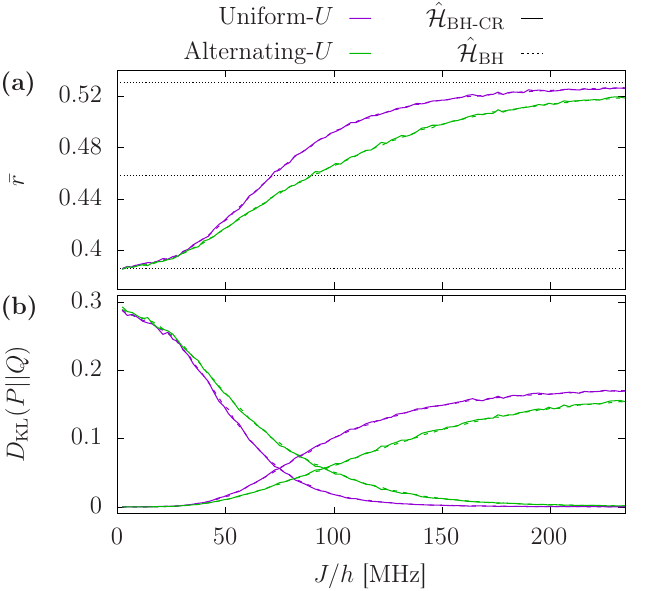}
    \caption{
    Influence of the counter-rotating terms on the level statistics. \textbf{(a)} Average level spacing ratio and \textbf{(b)} KL divergence 
    for uniform (purple), and alternating (green) BH arrays as functions of the hopping amplitude $J$, obtained including (solid) or neglecting (dashed) the counter-rotating (CR) terms. The dotted horizontal lines in panel (a) correspond to (top to bottom) $\bar{r}_1$, $(\bar{r}_0+\bar{r}_1)/2$, and $\bar{r}_0$. We consider 5000 disorder realizations for the dashed curves, where CR terms are neglected. For the solid curves, due to the larger Hilbert space (see text) we consider 1000 disorder realizations for $50< J/h< 150$~MHz, and 500 realizations for the remaining values of $J$. For all the data, we consider $\mathcal{N}=4$ particles and $M=10$ sites. The parameters used are as in \cref{tab:parameters} and mapped to the BH model as in \cref{fig:csfq}.
    }
    \label{fig:cr}
\end{figure}

\Cref{fig:cr} shows the average level spacing ratio and the KL divergence for the spectrum of a chain with $\mathcal{N}=4$, $M=10$ and averaged over disorder realizations (see caption), either including [solid, \cref{eq:bhm-cr}] or neglecting [dashed, \cref{eqn:H-BH}] the CR terms. The differences between the two cases for both the uniform (purple) and the alternating (green) systems are comparable to the statistical uncertainty in our dataset, showing that CR terms do not significantly affect the level statistics; this result also validates our perturbative approach, since the influence of higher order terms is expected to be even smaller.

\section{Level repulsion in a 3-site array}\label{app:3sites}

\begin{figure}
    \centering
    \includegraphics[width=\linewidth]{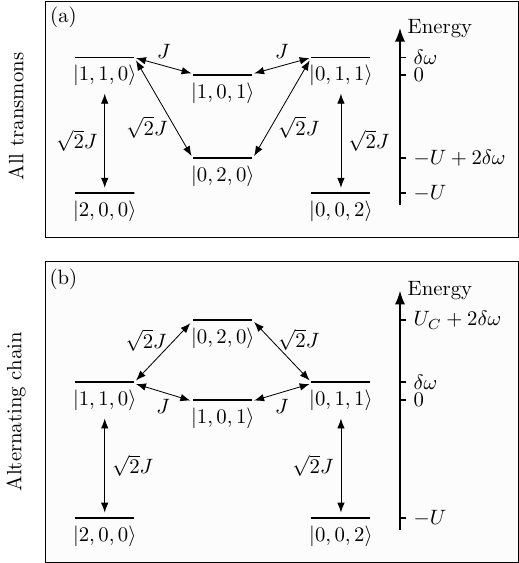}
    \caption{Level diagrams in a three-site ($M=3$) and two particle ($\mathcal{N}$=2) array with resonant frequency end-sites for (a) transmon and (b) alternating chains.
    The level repulsion between the states $\ket{2,0,0}$ and $\ket{0,0,2}$ depends on the energy of the state $\ket{0,2,0}$, and so on the anharmonicity of the central qubit. For small disorder, $\delta\omega\ll U$, arrays with alternating sign anharmonicity have a smaller level repulsion, thus favoring localization (see text).
    }
    \label{fig:lr}
\end{figure}

The enhanced resilience to chaos in alternating transmon-CSFQ chains compared to uniform chains for small disorder strengths $\hbar\delta\omega_{01}\ll U$ can be captured in a minimal setting, consisting of three sites ($M=3$) and two excitations ($\mathcal{N}=2$). For convenience, we set $\hbar=1$ hereafter. With no loss of generality, we consider arrays with transmons at the lateral sites (Q=T for $i=\{1,3\}$), and a generic qubit type in the central site $C$. To compare localization in systems with uniform and alternating-sign anharmonicity (interaction), we investigate the level repulsion between the second excited states in next-to-nearest sites, i.e., the states $\ket{2,0,0}$ and $\ket{0,0,2}$ in our minimal setting, where a suppressed level repulsion favors localization. More specifically, in a worst-case scenario for localization, we take the transmons to
have the same frequency ($\omega$) and anharmonicity ($-U$), 
so that
$\ket{2,0,0}$ and $\ket{0,0,2}$ are degenerate. Due to disorder, the qubit frequency in the central site is detuned by $\delta\omega$, and hence the state $\ket{0,2,0}$ is detuned from $\ket{2,0,0}$ and $\ket{0,0,2}$; specifically, the detuning amounts to $2\delta\omega$ and $U+U_C+2\delta\omega$ for all-transmon and alternating chains, respectively (see level diagrams in Fig.~\ref{fig:lr}). In fact, this difference determines the enhanced localization for alternating chains, as we detail here. The Hamiltonian of the disordered Bose-Hubbard model [Eq.~\eqref{eqn:H-BH}] for $\mathcal{N}=2$ and $M=3$ can be written in the basis $\{\ket{2,0,0},\ket{1,1,0},\ket{1,0,1},\ket{0,2,0},\ket{0,1,1},\ket{0,0,2}\}$ as the $6\times 6$ matrix, 
\begin{equation}
\hat{H}_{\rm BH} = 
  \begin{pmatrix}
    -U & \sqrt{2}J & 0& 0 & 0 & 0\\
    \sqrt{2}J & \delta\omega & J &\sqrt{2}J& 0 & 0\\
   0 & J & 0 & 0& J & 0\\
    0 &\sqrt{2}J & 0 & U_C+2\delta\omega& \sqrt{2}J & 0\\
    0 & 0 & J & \sqrt{2}J& \delta\omega & \sqrt{2}J\\
    0 & 0 & 0 & 0& \sqrt{2}J & -U
  \end{pmatrix}\, ,
  \label{eqn:BHM6x6}
\end{equation}
where we subtracted the irrelevant constant $2\omega$. The eigenvalues of the Hamiltonian in Eq.~\eqref{eqn:BHM6x6} can be computed explicitly for arbitrary values of the parameters $U$, $U_C$, $\delta\omega$, and $J$ since the characteristic polynomial of the matrix factorizes in a quadratic and a quartic equation. For our goals, it is sufficient to estimate the energy level splitting ($\Delta E$) of states $\ket{2,0,0}$ and $\ket{0,0,2}$ in the limit $J\ll U,\,|U+\delta\omega|,\,|U+U_C+2 \delta\omega|$. Using a standard perturbative approach for degenerate levels, the energy splitting at the leading order can be expressed as
\begin{equation}
\Delta E (\eta) \simeq \frac
{4J^4}{U^3(1+\delta\omega/U)^2}\left|1+\frac{2}{1+\eta+2\delta\omega/U}\right|\, ,
\label{eq:splitting}
\end{equation}
where $\eta=U_C/U$. Clearly, $\Delta E\propto J^4$ as four hopping processes are necessary to connect the states $\ket{2,0,0}$ and $\ket{0,0,2}$ (see Fig.~\ref{fig:lr}). The level splitting is obtained by summing over the two possible paths; both paths involve the states $\ket{1,1,0}$ and $\ket{0,1,1}$ and either the state $\ket{1,0,1}$ and $\ket{0,2,0}$. Each hopping between the states $j$ and $j'$ contributes a factor $V_{jj'}/(E_{j'}+U)$, where $V_{jj'}=\braket{j|\hat{H}_{\rm BH}|j'}$, $E_{j'}=\braket{j'|\hat{H}_{\rm BH}|j'}$, and $j'$ runs only over the intermediate states (that is, the denominator for the last hop is unity); clearly, the splitting for the path involving $\ket{0,2,0}$ depends on the central qubit type. Equation~\eqref{eq:splitting} captures the enhanced localization of an alternating chain (with $\eta>0$) over a uniform chain ($\eta=-1$); in fact at low disorder, $\delta\omega/U\ll 1$, the level splitting for the alternating chain in comparison to the uniform one is suppressed as
\begin{equation}
\frac{\Delta E(\eta)}{\Delta E (-1)}\simeq\frac{|\delta\omega|}{U}\frac{3+\eta}{1+\eta}\,,
\label{eq:suppressionDeltaE}
\end{equation}
and decreases monotonically with $\eta$ for a fixed disorder strength $\delta\omega$, consistently with the increased value of $J_C$ with $\eta$ [cf. Fig.~\ref{fig:disorder-scaling-alt}(b)]. In contrast, the relative splitting increases with $|\delta\omega|/U$, qualitatively in agreement with the reported decrease in $\Delta J_C/J_C$ with disorder, see Fig.~\ref{fig:disorder-scaling-alt}(a).

The result in Eq.~\eqref{eq:splitting} enables us to consider the effect of disorder in the anharmonicity on localization in uniform arrays. Let us take $\eta = -1 +\delta U/U$ with $\delta U \ll U$; in this case we find (still assuming $\delta\omega\ll U$)
\begin{equation}
 \frac{\Delta E(\eta)}{\Delta E (-1)}\simeq \left|1+\frac{\delta U}{2\delta\omega}\right|^{-1}  
\end{equation}
Therefore, for deviation $\delta U$ from uniform anharmonicity large compared to the frequency disorder $\delta\omega$, we expect localization to persist to stronger hopping $J$ than in the case of nearly-uniform anharmonicity $\delta U \ll \delta\omega$.

\bibliography{bib}

\end{document}